\documentclass[twocolumn,amsmath,amssymb,aps]{revtex4-2}

\usepackage{graphicx}
\usepackage{dcolumn}
\usepackage{bm}

\usepackage{ulem}

\newcommand{\jgr}{{\it J. Geophys. Res.}}
\newcommand{\planss}{{\it Planet. Space Sci.}}
\newcommand{\ssr}{{Space~Sci.~Rev.}}

\newcommand{\solphys}{{\it Solar Phys.}}
\newcommand{\noopsort}[1]{}
\newcommand{\apjl}{ApJ Letters}
\newcommand{\apjs}{ApJ Supplement}

\newcommand{\mnras}{{\it Mon. Not. Roy. Astron. Soc.}}
\def\physrep{\ref@jnl{Phys.~Rep.}}   

\begin{document}
\title{The Interaction between Shocks and Plasma Turbulence: Phase Space Transport}

\author{D. Trotta$^1$, F. Valentini$^1$, D. Burgess$^2$, and S. Servidio$^1$}
\affiliation{
$^1$Dipartimento di Fisica, Universit\`a della Calabria, I-87036 Cosenza, Italy\\
$^2$School of Physics and Astronomy, Queen Mary University of London, Mile End Road, London E1 4NS, United Kingdom}

\date{\today} 


\date{\today}

\begin{abstract}

The interaction of collisionless shocks with fully developed plasma turbulence is numerically investigated. Hybrid kinetic simulations, where a turbulent jet is slammed against an oblique shock, are employed to address the role of upstream turbulence on plasma transport. A novel technique, using coarse-graining of the Vlasov equation, is proposed, showing that the transport properties strongly depend on upstream turbulence strength, with turbulent structures-modulated patterns. These results might be relevant for the understanding of acceleration and heating processes in space plasmas.

\end{abstract}


\maketitle


A turbulent plasma wind flows from the Sun and permeates the Heliosphere, encountering several magnetic obstacles, leading to shocks that continuously interact with the incoming complex solar wind -- a scenario that becomes a prototype for understanding many other systems characterised by the presence of shocks. Despite decades of research, the interaction of shocks with plasma turbulence and the subserquent energetic particle production still remain poorly understood \citep{Bykov2019,Lazarian2012}. Turbulence-generated coherent structures and waves might interact with the shock discontinuity, in an interplay which is likely to play a pivotal role in particle acceleration and plasma heating \citep{Kennel1985,Zank2002,Giacalone2008}.

On very large scales, shocks are responsible for the formation of giant radio relics, i.e., elongated structures showing strong, polarised radio emission at the interface between clusters of galaxies \citep{Brunetti2014}. Shocks are well-known efficient, natural particle accelerators \citep{Burgess2015} and have been modelled in a number of theories \citep{Axford1977,Bell1978a,Krymskii1977,Blandford1978, Katou2019}. Less understood is the interaction of shocks and turbulence, that characterizes spectacular high energy events, as in supernovae explosions propagating through the interstellar turbulent medium, as in the case of coronal mass ejections that stream through the turbulent solar wind, as for the complex Earth's bow shock environment. In many of the above examples, oblique shocks are known to generate coherent Field Aligned Beams (FABs), as observed at Earth's bow shock~\citep{1Paschmann1980}. FABs are an important source of free energy throughout the interplanetary medium ~\citep{Kucharek2004}.

Turbulence is populated by a variety of structures that can work effectively as particle ``traps'' and ``corridors'', that either hinder or enable their motion \citep{Trotta2020b}, and represents another crucial source of accelerated particles \citep{Jokipii1966,Dmitruk2004,Drake2006, Comisso2019}. An example of such energization process has been observed in the patterns of local reconnection that develop in turbulence \citep{Servidio2009, Pecora2019}. In order to understand such mechanisms, the transport properties need to be explored in the plasma phase space~\citep{Parker1965}.

Due to the difference between the spatial and temporal scales involved in accelerating particles, shocks and turbulence are often considered on parallel theoretical paths. However, fundamental studies have suggested that these are inextricably linked: shocks are likely to propagate in turbulent media, and turbulence is responsible for changing fundamental aspects of shock transitions \citep[][]{Jackson1993,Giacalone2005a,Giacalone2005b,Guo2015,Caprioli2014b,Haggerty2020}. Inspired by these studies, here we quantitatively explore the intimate relation between these two phenomena.

Hybrid Vlasov-Maxwell PIC simulations have been employed in 2D, where a turbulent jet, generated via compressible MagnetoHydroDynamics (MHD) simulations, has been shot against a supercritical shock. We investigate the shock-turbulence interaction and its role on the particles transport, using both Lagrangian and Eulerian approaches. We present a novel technique to investigate the transport processes at play, based on coarse-grained techniques, typical of fluid-dynamics, accompanied by velocity space integration, typical of energetic particle transport models.

\begin{figure}
\includegraphics[width=0.48\textwidth]{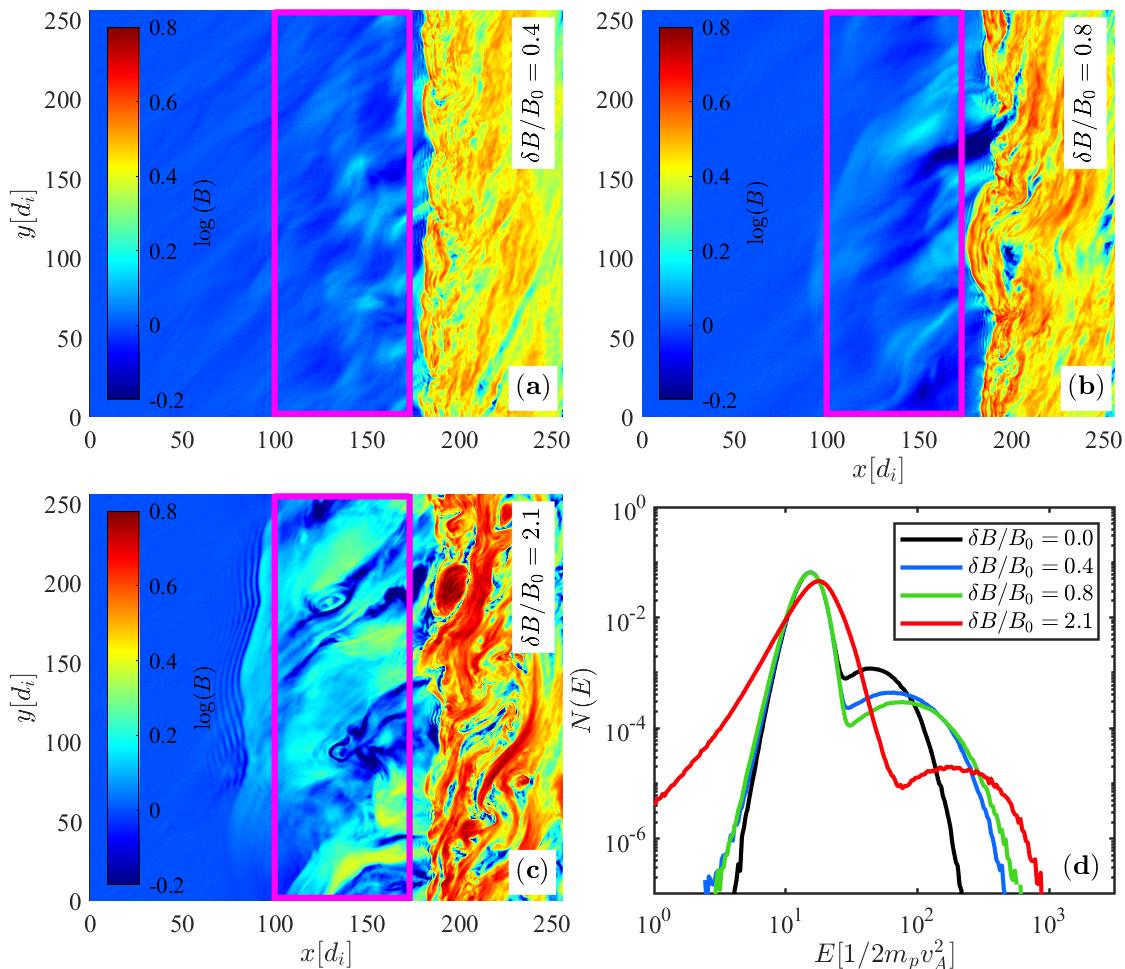}
\caption{(a) 2D colormaps of magnetic field magnitude $B$ for the perturbed shocks and with different upstream turbulence strengths (a)-(c). (d) Upstream energy spectra, collected in the regions highlighted by the magenta boxes.}
\label{fig:fig1}
\end{figure}

The methodology is based on two stages. First, fully developed, decaying turbulence is generated by means of 2.5D MHD simulations \citep{Perri2017}. Four simulations were performed, with different levels of turbulence fluctuations, namely $\delta B/B_0 = 0.0, 0.4, 0.8, 2.1$, where $B_0$ is the mean field, at $\theta_{Bn} = 45^\circ$ in the $x$-$y$ plane, and $\delta B$ is the level of fluctuations. When the turbulence is fully developed and coherent structures form, the second (main) step consists of using the MHD output, with adequate windowing, as an upstream condition for a kinetic, hybrid Particle-In-Cell (PIC) simulation of a supercritical shock performed through the HYPSI code~\citep{Trotta2019}. In the main stage, the Vlasov-Maxwell equations are solved with fluid electrons and kinetic ions, and the  injection method is used~\citep{Quest1985}, where the shock propagates in the negative $x$-direction, while the turbulent pattern moves oppositely, in the upstream region.

In the MHD simulations, typical Alfv\'en units have been used, on a periodic box. In the hybrid PIC simulations, distances are normalised to the ion inertial length $d_i \equiv c/\omega_{pi}$, times to the inverse cyclotron frequency ${\Omega_{ci}}^{-1}$, velocity to the Alfv\'en speed $v_A$ (all referred to the unperturbed upstream state), and the magnetic field and density to their unperturbed upstream values ($B_0$ and $n_0$). An upstream flow with $V_\mathrm{in} = 3.5 v_A$  has been chosen, resulting in an Alfv\'enic Mach number of the shock $M_A \simeq 5.5$. The upstream ion distribution function is an isotropic Maxwellian and the ion $\beta_i = 1$, as in the MHD simulation.  The simulation domain  is 256 $d_i$ $\times$ 256 $d_i$, with a grid size $\Delta x$ = $\Delta y$ = 0.5 $d_i$ and a particle time-step $\Delta t_{pa}$  = 0.01 $\Omega_{ci}^{-1}$. The number of particles per cell used is large, always greater than 500 (upstream), in order to keep the statistical noise at a very low level. 

This novel technique represents a realistic step-forward with respect to ``laminar'' injection and is different from other perturbation methods, where uncorrelated random noise or a prescribed spectrum of fluctuations can be added to the upstream \cite{Giacalone2005b, Guo2015}. Here, turbulence consists of a fully developed spectrum of fluctuations, with a large variety of coherent structures and waves which are crucial for the transport properties, as predicted by important theoretical works \citep{Zank2002, LeRoux2018}, as will be discussed.

\begin{figure}
\includegraphics[width=0.45\textwidth]{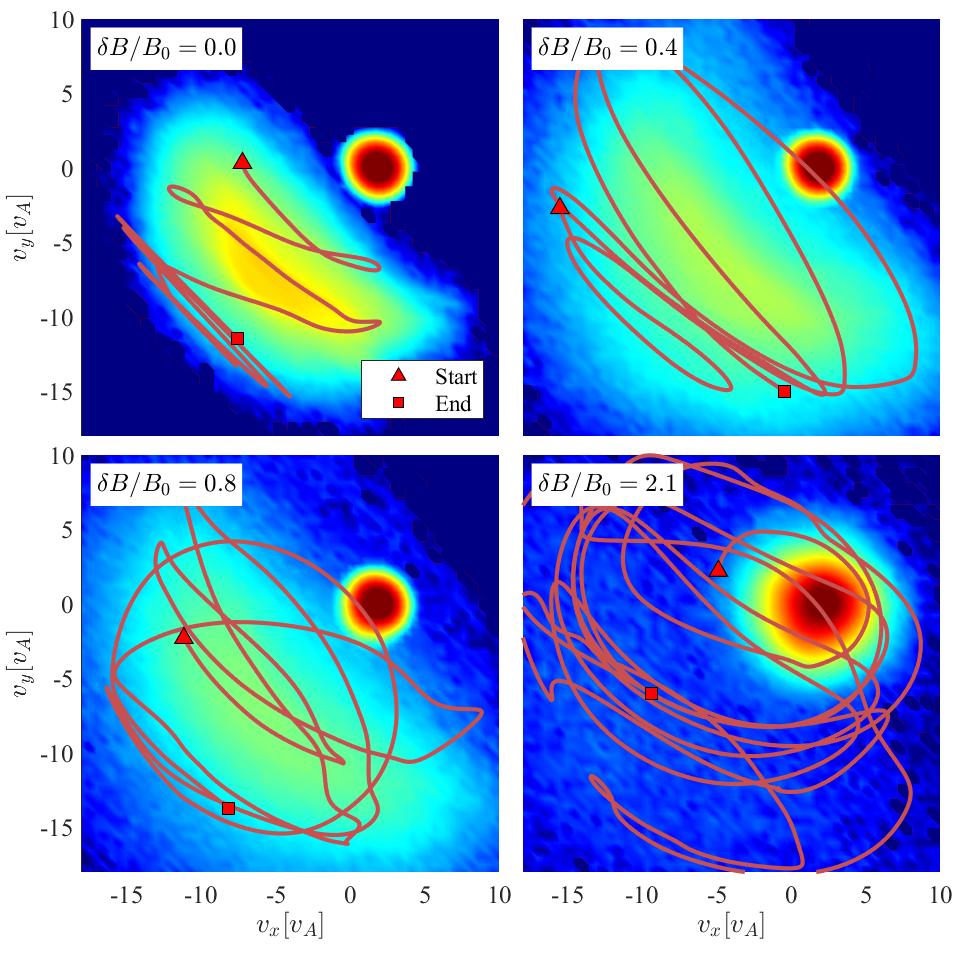}
\caption{Protons VDFs in the shock upstream (collected in the magenta boxes in Fig. \ref{fig:fig1}), at different turbulence levels. Particles trajectories are reported from the beginning to the end of the shock-turbulence interaction.}
\label{fig:fig2}
\end{figure}

Fig.~\ref{fig:fig1} (panels a-c) shows an overview of the perturbed simulations, where we report the 2D colormaps of the magnetic field intensity $B$. The shock front is interacting with the perturbed upstream and there is a net change of topology for increasing turbulence level. Panel (d) shows the upstream energy spectra, for all cases. When $\delta B/B_0=0$ (unperturbed case), the spectrum shows the Maxwellian inflow population together with a narrow beam of accelerated particles, namely the FAB. Turbulence manifests in three ways. First, for higher upstream turbulence strength, particles achieve higher energies. Second, the high energy FAB shifts and spreads for increasing $\delta B/B_0$, suggesting that some mechanism of beam ``decoherence'' is at play.  The third effect is the production of very low energy particles, evidently related to a process of particle deceleration and trapping. These features are possibly due to field-particle interactions, where typical turbulence patterns act as spreaders or transport barriers ~\citep{Servidio2016, Pecora2019, Trotta2020b}.  These changes in energy spectra are intimately related to phase space transport and diffusion, as discussed later. 

In order to extract more details about the transport processes, we reconstruct the upstream particles Velocity Distribution Functions (VDFs), fundamental for an Eulerian approach \citep{Gleeson1969,Jokipii1986a,Zank2015, LeRoux2018}. In Fig.~\ref{fig:fig2} we report examples of such VDFs, integrated along $v_z$, in the  $v_x-v_y$ velocity space, for all the simulations. In the unperturbed case, the inflow population and the reflected FAB are well-separated. When upstream turbulence is present, the separation between the two populations is much less sharper. A ``distortion'' of the inflow population is observed, due to the turbulent plasma heating at the shock front, particularly prominent for the most turbulent case. Following in sequence the perturbation amplitudes, it is evident that turbulence smooths and diffuses the two particle populations, that spread toward both high and lower energies, as reported in Fig. \ref{fig:fig1}-(d). It is now natural to ask how particles behave in the velocity subspace. In Fig.~\ref{fig:fig2}, some typical phase space trajectories of energetic protons, superimposed on the VDFs, are reported. While in low turbulence cases the particles remain confined in sectors of the velocity space, they can break down transport barriers thanks to higher turbulence levels. On this observation we base our new method, described below.

\begin{figure}
\includegraphics[width=0.98\columnwidth]{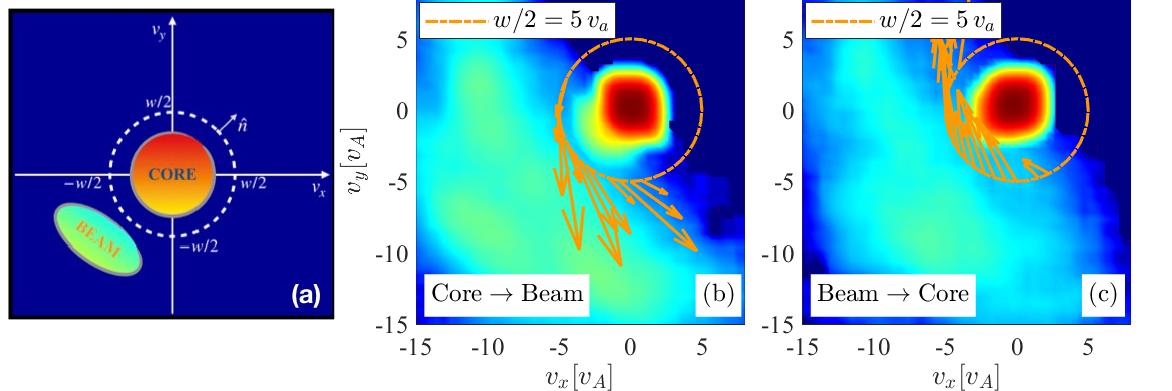}
\caption{(a) Sketch of typical particle distributions together with the integration circle $\gamma_w$ (dashed-white). (b) Upstream VDF in coarse-grained cell with positive flux $\int_\gamma  \overline{F}_l \overline{\mathbf{P}}_l \cdot \hat{\mathbf{n}} \,  d \gamma $. The vector $\overline{F}_l \overline{\mathbf{P}}_l $ is shown on the integration path (brown arrows). (c): same as (b) but in a region with negative flux. \label{fig:fig3}}
\end{figure}

The premise is that the shock-turbulence interaction is a multi-scale process characterized by a variety of ``lengths'', that in a plasma involve both the physical and velocity sub-space. In such multidimensional complexity, it is natural to approach via coarse-graining and filtering techniques, successfully used, for decades, in fluid dynamics \citep{Leonard1974}. Recent progresses have been made also in plasma kinetic equations \citep{Eyink2018}, as well as in fluid plasma modeling \citep{Camporeale2018}. 

The scope is to characterize the phase space in sectors where dispersion and trapping, acceleration and deceleration might compete. Our novel technique relies on a coarse-graining of the Vlasov equation \cite{Eyink2018}, combined with a computation of the reduced (coarse-grained) moments. In order to simplify our model at a basic level, we reduce to a 2D-2V description. By integrating along $v_z$, we define $F(x,y,v_x,v_y,t) = \int_{-\infty}^{+\infty}f(\mathbf{x},\mathbf{v},t)  \mathrm{d} v_z$, where $f(\mathbf{x}, \mathbf{v},t)$ is the ion VDF. This leads to 
\begin{equation}
	\label{eq:Vlasov2}
    \partial_t F + \nabla \cdot \left( \mathbf{v} F \right) + \nabla_v \cdot \left( \mathbf{P} F \right) = 0, 
\end{equation}
where ${\bf E}$ and ${\bf B}$ are the electric and magnetic fields, respectively, and the acceleration is given by $\mathbf{P}({\bf x}, {\bf v})={\bf E}+{\bf v}\times{\bf B}$, on reduced 4D phase space with coordinates ($x, y, v_x, v_y$). We concentrate on a coarse-grained Vlasov equation, by defining a scale-dependent, filtered distribution
\begin{equation}
    \overline{F}_l(\mathbf{x}, \mathbf{v}, t) = \int_{-\infty}^{+\infty} f(\mathbf{x} + \mathbf{r}, \mathbf{v}, t) G_l(\mathbf{r}) d^2 r, 
\end{equation}
where $G_l(\mathbf{r})$ is a kernel that satisfies a series of properties, being non-negative, normalized, centered and rapidly approaching to zero for $r\rightarrow \infty$. In our case we chose the simplest box-filter type, that in the reduced 2D Cartesian coordinate is given by $ G_l(\mathbf{r}) = 1/l^2$ for $|r_x|<l/2$ and  $|r_y|<l/2$, and equal to zero otherwise. 
By filtering Eq.~(\ref{eq:Vlasov2}) is easy to get
\begin{equation}  
\label{eq:CGV1}
    \partial_t \overline{F}_l + \nabla_l \cdot \left[ \mathbf{v} \overline{F}_l \right] + \nabla_v \cdot \left[ \overline{\mathbf{P}}_l \overline{F}_l \right] = \nabla_v \cdot \overline{\mathbf{Q}}_l,
\end{equation}
where $\overline{\mathbf{P}F}_l = \overline{\mathbf{P}}_l \overline{F}_l - \overline{\mathbf{Q}}_l$. The latter decomposition, typical of Reynolds-averaging techniques \citep{Germano1991}, introduces the ``closure problem'', related to the description of the subgrid modeling \citep{Yang2018}, equivalent to turbulent diffusion due to small scale eddies.

In analogy with the Parker equation for energetic particles transport, we choose a typical speed $w$ (and therefore a typical energy) at which we integrate Eq.~\ref{eq:CGV1} \citep{Jokipii1987}. The zeroth-reduced moment \citep{Parker1965,Jokipii2010}, namely the reduced particle density becomes
\begin{equation}
\overline{N}_{l, w}(x,y,t) = \int d^2 v \overline{F}_l(x, y, {\bf v}, t) G_w({\bf v}),
\end{equation}
where now $G_w({\bf v})$ is a unitary kernel, different from zero only for $|{\bf v}|<w$ and $\overline{N}_{l, w}(x,y,t)$ is the thermal population, coarse-grained in space at scale $l$. Applying $G_w({\bf v})$ and integrating Eq.~\ref{eq:CGV1}, one gets
\begin{equation}
	\label{eq:CGV2}
    \partial_t \overline{N}_{l,w} + \nabla_l \cdot\left[ \overline{N}_{l,w} \overline{\mathbf{V}}_{l,w}  \right] 
    + \int_{\gamma_w}\!\!\!\!\overline{F}_l \overline{\mathbf{P}}_l \cdot \hat{\mathbf{n}}\,d\gamma 
    = \int_{\gamma_w}\!\!\!\!\overline{\mathbf{Q}}_l \cdot \hat{\mathbf{n}}\,d\gamma.
\end{equation}
The first term represents the time variation of the reduced density, the second term is responsible for spatial transport of particles over the coarse-grained space, while the third represents the flux across a circle $\gamma_w$ of radius $w$. The right-hand side is the residual contribution from the subgrid scales. Interestingly, the velocity-space propagation term of Eq.~(\ref{eq:CGV2}) is essentially due to the normal component of the electric field, while the magnetic part of $\overline{\mathbf{P}}_l$ is tangent to the surface, acting as a pitch-angle spreader \citep{Lyons1974}. Eq.~(\ref{eq:CGV2}) results from the 2D divergence theorem in velocity space for plasmas, for spatial integration at a length-scale $l$ and at a velocity cutoff $w$. The advantage of the model is to describe the large scale patterns of acceleration and diffusion processes. A simple study of the sign of the fluxes provide unique information about spatial diffusion/clusterization and about energization/deceleration.

\begin{figure}[ht]
\includegraphics[width=.45\textwidth]{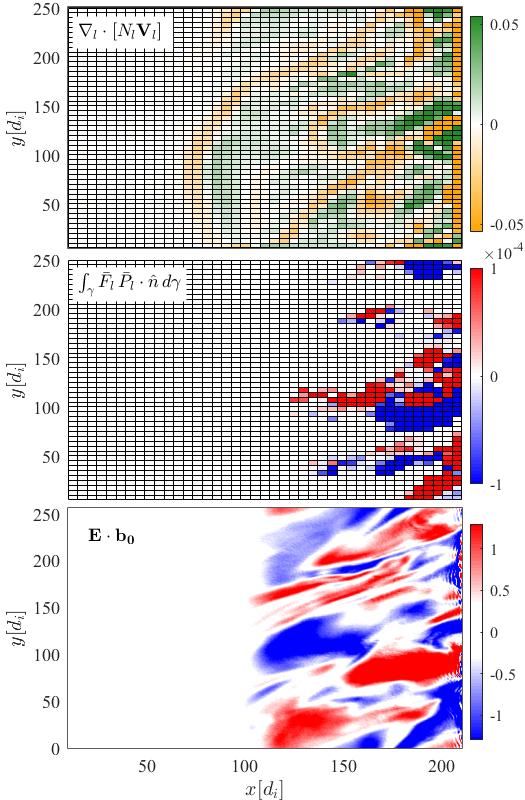}
\caption{Coarse grained mosaic of the turbulence-shock interaction. Upstream spatial (top) and velocity (bottom) transport terms, for moderate turbulence level, for coarse-graining at $l=5 d_i$ and $w = 10 v_A$. Bottom: parallel electric field.  \label{fig:fig4}}
\end{figure}

An example of this Eulerian approach to velocity space diffusion is shown in Fig.~\ref{fig:fig3}-(a), with a cartoon of our velocity space integration circle $\gamma_w$ and its normal $\hat{\mathbf{n}}$. The spatial coarse-graining $l$ was chosen to be of 5 $d_i$, a value in the inertial range of the upstream turbulence spectrum. This is very important since the results obtained in this range are self-similar (not shown here), typical of inertial range coarse-graining in fluids \citep{Lesieur1996}. We chose the $w$ parameter to be of 10 $v_A$, in such a way that the integration in velocity space is done between the core and the beam population. In Fig.~\ref{fig:fig3}-(b) and (c) we represent $\overline{F}_l({\bf x}, {\bf v}, t)$, at two different spatial cells, with positive (b) and negative (c) flux-integral $\int_{\gamma_w}\!\!\!\!\overline{F}_l \overline{\mathbf{P}}_l \cdot \hat{\mathbf{n}}\,d\gamma$. The VDFs are obtained from the intermediate turbulence case, $\delta B/B_0 = 0.8$, typical of solar wind conditions. A positive net flux through $\gamma_w$ indicates an energization mechanism, where $\overline{N}_{l,w}$ diminishes. In the opposite case, a deceleration or cooling mechanism is at work and the core gains particles (Fig.~\ref{fig:fig3}-(c)).

How do these patterns look like in space? Following our governing Eq.~(\ref{eq:CGV2}), we can identify regions of strong spatial dilatation (compression) and regions of strong acceleration (deceleration). In Fig.~\ref{fig:fig4} we report this characterization of the phase-space transport, over a ``pixelized'' domain with resolution $l=5 d_i$. The top panel shows a spatial transport overview. When the space-transport is positive, thermal particle are escaping, evidently subject to the Alfv\`enic turbulence. On the other hand, when the spatial transport term is negative, plasma condensates. The overall picture for upstream velocity-space transport is shown in the middle panel of Fig.~\ref{fig:fig4}, being enhanced at the turbulence-shock boundary layer. As in the mechanism described in Fig.~\ref{fig:fig3}, a positive velocity transport ($\int_{\gamma_w}\!\!\!\!\overline{F}_l \overline{\mathbf{P}}_l \cdot \hat{\mathbf{n}}\,d\gamma$) indicates a net flux of the core population towards higher energies (beam population), and vice-versa for very negative fluxes.

\begin{figure}
\includegraphics[width=.45\textwidth]{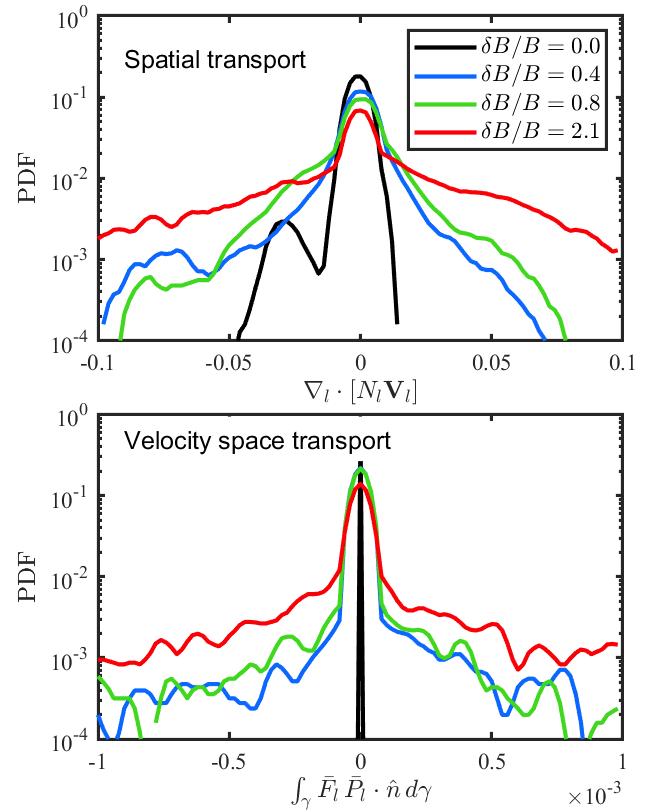}
\caption{PDFs of the spatial (top) and velocity-space (bottom) fluxes, for different upstream turbulence cases. The turbulence-shock interaction enhances the transport.\label{fig:fig5}}
\end{figure}

The enhanced velocity-fluxes are evidently due to the interplay between particles and turbulent fields, where the ${\bf v}\times{\bf B}$ force acts as a pitch angle spreader along the surface and the turbulent electric field locally enhances momentum diffusion, via local processes such as wave-particles interaction, linear and non-linear Landau damping \citep{Howes2017, Chen2019}, stochastic ion heating \citep{Chandran2013} and possible interaction with reconnetion processes in the upstream turbulent layer \citep{Zank2015, Servidio2009, Howes2017}. In order to establish such field-particle synergy and quantify the net transport across $\gamma_w$, we computed the electric field parallel to the mean field ${\bf B}_0$. As it can be seen from Fig.~\ref{fig:fig4}-(c), there is a very good correlation between $E_{||}={\bf E} \cdot {\hat{\bf b}}_0$ and the velocity transport, especially for the most extreme values. We evaluated the correlation coefficient between the two terms, finding $C\simeq 0.7$. Large fluxes and parallel electric field are, as expected, anti-correlated, suggesting that any positive parallel electric field energize particles.

In Fig.~\ref{fig:fig5} we compare all the numerical experiments, by computing the probability density functions (PDFs) for both spatial and velocity space transport terms. Increasing upstream turbulence enhances the phase space transport, thus  explaining the broader energies observed in Fig.~\ref{fig:fig1}.  In the unperturbed case, the velocity space transport is very small: the core and beam population appear well-separated when upstream perturbations are not present. In this scenario, particle acceleration happens only at the shock front. The efficiency of the mixing depends dramatically on the turbulence level, although even a small amount of turbulence is a very efficient diffusor.

In summary, the plasma behavior upstream of oblique shocks has been investigated in the presence of pre-existing, MHD-generated turbulence. A dramatic change of the plasma transport has been found, going from the unperturbed to the super-turbulent case (Fig.~\ref{fig:fig1}), which has been investigated by using both Lagrangian and Eulerian approaches. From the Lagrangian point of view, particles can escape from their original population thanks to a ``bridge'' established by turbulence (Fig.~\ref{fig:fig2}). In order to understand such behavior, a novel Eulerian technique, based on the coarse-graining of the Vlasov equation, has been proposed, where we combine spatial filtering, typical of hydrodynamics, with Parker-type transport equations, typical of cosmic ray physics \citep{Amato2017}. As it can be seen from Fig.~\ref{fig:fig4}, by averaging over inertial range scales and by using a divergence theorem in velocity space, the turbulent upstream is made up of a ``mosaic'', where each piece of such puzzle is characterized from strong spatial dilation and condensation, thanks to the Alfv\`enic turbulent modulations. More interestingly, the $v$-space filtered flux is noticeably anti-correlated with parallel electric field, suggesting the possibility of several field-particles interactions. 

These results might have important consequences on the understanding of transport and heating processes in a variety of systems, ranging from the Earth's bow shock interacting with the turbulent solar wind, to the largest scales of radio relics in galaxy clusters \citep{Ha2018} . In future works we will extend the analysis to the full phase-space and including the effect of kinetic electrons, with possible applications to \emph{in-situ} measurements \citep{Servidio2017, Pezzi2018}.

\begin{acknowledgments}
This work has received funding from the European Unions Horizon 2020 research and innovation programme under grant agreement No. 776262 (AIDA, www.aida-space.eu).
\end{acknowledgments}

\end{document}